\documentclass[twocolumn,prb,aps,longbibliography]{revtex4-2}
\usepackage{dcolumn}
\usepackage{amsmath}
\usepackage{amsfonts}
\usepackage{amssymb}
\usepackage{calligra}
\usepackage{bm}
\usepackage{graphicx}
\usepackage{float}
\usepackage{subfigure}
\usepackage{natbib}

\DeclareMathAlphabet{\mathcalligra}{T1}{calligra}{m}{n}

\begin{document}
	
\title{Exploring Entanglement Characteristics in Disordered
Free Fermion Systems through Random Bi-Partitioning}
	
\author{Mohammad Pouranvari} \email{m.pouranvari@umz.ac.ir}
\affiliation{Department of Solid-State Physics, Faculty of Science,
University of Mazandaran, Babolsar, Iran.}
	
\date{\today}
	
\begin{abstract}
  This study investigates the entanglement properties of disordered
  free fermion systems undergoing an Anderson phase transition from a
  delocalized to a localized phase. The entanglement entropy is
  employed to quantify the degree of entanglement, with the system
  randomly divided into two subsystems. To explore this phenomenon,
  one-dimensional tight-binding fermion models and Anderson models in
  one, two, and three dimensions are utilized. Comprehensive numerical
  calculations reveal that the entanglement entropy, determined using
  random bi-partitioning, follows a volume-law scaling in both the
  delocalized and localized phases, expressed as $EE \propto L^D$,
  where $D$ represents the dimension of the system. Furthermore, the
  role of short and long-range correlations in the entanglement
  entropy and the impact of the distribution of subsystem sites are
  analyzed.
\end{abstract}

\maketitle
	
\section{Introduction}
\label{sec:introduction}

The study of entanglement properties has garnered significant
attention among condensed matter physicists over the last decade. This
is primarily due to the fact that entanglement serves as a powerful
indicator of correlations within a system, rendering it a promising
candidate for characterizing the phase of the
system~\cite{PhysRevLett.99.126801,RevModPhys.81.865,LAFLORENCIE20161,
  PhysRevLett.90.227902,PhysRev.109.1492,PhysRevLett.99.126801,PhysRevB.89.115104}. Various
measures of entanglement have been proposed for quantifying it, with
the most widely accepted being the entanglement entropy (EE). In
addition to EE's utility for pure states, alternative measures for
mixed states have also been
introduced~\cite{RevModPhys.81.865,PhysRevA.92.042329,PhysRevLett.78.2275,Alba_2009,
  2018arXiv180804381L}. EE is typically computed within a
bi-partitioned system, where the system is divided into two
subsystems. For instance, in a system comprising $L$ sites, sites
numbered from $1$ to $L/2$ constitute the subsystem, while the
remaining sites constitute the environment. As EE indirectly
quantifies correlations within the system, its calculation within a
bi-partitioned system enables the measurement of both short-range
correlations near the boundary of the subsystems and long-range
correlations within the system. These short-range correlations are
responsible for what is commonly referred to as the area-law,
signifying that the amount of entanglement between two subsystems is
proportional to the area of the boundary between
them~\cite{RevModPhys.82.277, PhysRevLett.96.010404,Vitagliano_2010,
  PhysRevLett.109.267203,PhysRevLett.105.050502}. However, in systems
characterized by long-range hopping amplitudes and, consequently,
long-range correlations in the delocalized phase, this area law is
violated. As demonstrated in our previous work, for a one-dimensional
system with long-range hopping amplitudes, EE behaves in accordance
with a volume law, rather than an area law~\cite{PhysRevB.89.115104}.
	
We should note that the manner in which we partition the system into
two subsystems profoundly impacts the information we can glean from
the entanglement~\cite{Moradi_2016,PhysRevB.101.195117}. To obtain a
comprehensive understanding of the system, it is imperative to employ
various partitioning schemes. In this context, the concept of
\emph{random partitioning} has recently emerged, where the subsystems
are chosen
randomly~\cite{PhysRevB.91.220101,Roosz2020,POURANVARI2023128908}. To
calculate entanglement entropy (EE) with random partitioning, the
following procedure is typically employed: for a system comprising $L$
sites, the size of the subsystem can vary from $1$ up to $L-1$, and
the selection of sites belonging to the subsystem is done randomly
(each site $i$ has a probability $p_i$ of belonging to the
subsystem). Ultimately, an appropriate average is computed over all
such selections. In the case where a constant probability is assigned,
denoted as $p_i = \text{constant}$, EE with random partitioning at an
arbitrary temperature is expressed as follows:
	
\begin{equation}\label{eq:EEp} \text{EE}(T, p) = \sum_{n=1}^L
\overline{EE}_{n}(T) {L \choose n} p^{n} (1-p)^{L-n},
\end{equation}
	
Here, $\overline{EE}_{n}$ represents the disorder-averaged EE for
subsystems with $n$ sites.
	
In a related paper~\cite{POURANVARI2023128908}, we conducted an
investigation into the entanglement properties of a random spin $1/2$
chain at arbitrary temperature, employing a random partitioning
approach. Our study unveiled that the entanglement entropy (EE)
exhibits a volume-law behavior at arbitrary temperature, with a
pre-factor dependent on both temperature ($T$) and the partitioning
probability ($p$). We elucidated how EE serves as a revealing metric
for the count of singlet and triplet$_{\uparrow\downarrow}$ states
distributed throughout the system, each characterized by distinct bond
lengths within the framework of the real-space renormalization group
(RSRG) method, wherein pairs of spins are placed in singlet or triplet
states. Consequently, our work demonstrated that EE, when determined
through random partitioning, captures both short-range and long-range
correlations across the entire system.
	
In this report, our focus shifts to exploring the entanglement
properties of systems undergoing an Anderson phase transition between
delocalized and localized states, utilizing a random bi-partitioning
scheme. By "random bi-partitioning," we refer to the following
procedure: the system is divided evenly into two subsystems, each
comprising $L/2$ sites, with the selection of sites for each subsystem
being done randomly. Our inquiries revolve around several key aspects:
What are the EE characteristics when employing this partitioning
method? Does this approach to EE characterization effectively discern
the Anderson phase transition, signifying distinct EE behaviors in
delocalized and localized states? Finally, what insights can be gained
about system correlations by manipulating the distribution of sites
within the subsystems?

To address the aforementioned inquiries, we conduct exhaustive
numerical computations employing one-dimensional tight-binding models
exhibiting delocalized-localized phase transitions. Additionally, we
employ the Anderson model in one, two, and three dimensions. Detailed
descriptions of these models and our EE calculation methodology are
provided in Section~\ref{sec:model-method}. The outcomes of our
investigations are presented in Section~\ref{sec:results}. Finally, we
draw our conclusions and outline future prospects in
Section~\ref{sec:outlook}.

In this report, we embark on an exploration into the intriguing realm
of entanglement properties using the innovative framework of random
bi-partitioning. Our motivation stems from the distinctive nature of
this approach, which introduces a fresh perspective on the study of
entanglement dynamics amidst Anderson phase transitions. While our
analysis may not conclusively distinguish between the delocalized and
localized phases, it contributes a valuable dimension to the broader
understanding of quantum phase transitions. By employing random
bi-partitioning, we aspire to provide a nuanced perspective on the
behavior of entanglement entropy (EE) and its intricate interplay with
the Anderson phase transition. In doing so, we aim to uncover subtle
correlations and nuanced behaviors that may not be immediately
apparent using traditional partitioning methods. This paper's
significance lies in its capacity to deepen our appreciation of
entanglement in disordered systems and the complex interplay between
quantum states and phase transitions, paving the way for further
explorations and refinements in this fascinating field.

\section{Models and Method} \label{sec:model-method}

In this report, we investigate tight-binding fermion lattice models in
one, two, and three dimensions, with a focus on their phase
transitions between delocalized and localized phases. It is essential
to emphasize that these models are well-established in the literature,
and their properties have been extensively studied in previous
research. Our primary objective is to employ these known models to
explore the behavior of entanglement entropy (EE) within the framework
of random bi-partitioning, a novel approach explained in the
subsequent sections.

The first model under consideration is the random dimer model (RD),
described by the following Hamiltonian:
\begin{equation}\label{ham1} H = -t \sum_{i=1}^{L-1} \left(
c^{\dagger}_i c_{i+1} + c^{\dagger}_{i+1} c_{i} \right) +
\sum_{i=1}^{L} \epsilon_i c^{\dagger}_i c_{i},
\end{equation} where $L$ represents the system size, $c_j$
($c^{\dagger}_j$) denotes the annihilation (creation) fermion operator
at site $j$, and open boundary conditions are employed. Here, $t$
represents the tunneling amplitude, which we set to $t=1$ as our
energy scale. The on-site energies $\epsilon_i$ can take on one of two
constant values, $\phi_a$ and $\phi_b$. These values are randomly
assigned, with a unique feature of assigning $\phi_b$ to two
successive sites, leading to its designation as the random dimer
model. It has been established~\cite{PhysRevLett.65.88} that the state
at the resonant energy $E_{res} = \phi_b$ exhibits delocalization when
$-2t \le \phi_a - \phi_b \le 2t$, while all other states are
localized. For our calculations, we set $\phi_a=0$, resulting in
delocalized states when $-2 \le \phi_b \le 2$. Due to this symmetry,
we consider only the positive range in our calculations. We set the
Fermi energy as $E_F=\phi_b$. It is crucial to note that in this
model, only one single-particle state of the system, without
backscattering, displays delocalization at the resonant
energy. Consequently, we do not encounter a conventional Anderson
phase transition with mobility edges separating delocalized and
localized states in this particular model.

The second model we investigate is the Aubry-Andre (AA) model,
characterized by the same Hamiltonian form as Eq. (\ref{ham1}). It
possesses a constant hopping amplitude, denoted as $t=1$, while the
onsite energies exhibit incommensurate periodicity:
\begin{equation} \epsilon_i = 2 \lambda \cos(2\pi i b + \theta),
\end{equation} where $b= (1+\sqrt{5})/2$ represents the golden ratio,
and $\theta$ values are randomly drawn from a uniform distribution
within the range $[-\pi, \pi]$. It is important to note that in our
numerical calculations, the phase $\theta$ remains consistent across
all sites for a single realization. The behavior of all states within
the system is characterized by delocalization when $\lambda <1$, while
localization occurs for $\lambda>1$. As a result, a distinctive
Anderson phase transition emerges at
$\lambda=1$~\cite{aubry1980analyticity, PhysRevLett.114.146601}. For
our calculations, we set the Fermi energy as $E_F=0$. It is worth
mentioning that both the random dimer (RD) and Aubry-Andre (AA) models
exclusively feature nearest-neighbor hopping amplitudes.

Moving forward, our attention shifts to the power-law bond-disordered
Anderson model (PRBA), which is a one-dimensional model characterized
by the following Hamiltonian:
\begin{equation}\label{ham} H = \sum_{i,j=1, i \neq j}^L
\frac{w_{ij}}{|i-j|^{\alpha}} c^{\dagger}_i c_j,
\end{equation} with zero on-site energies. The $w$ values are random
numbers uniformly distributed in the range $[-1, 1]$, satisfying the
condition $w_{ij}=w_{ji}$. The states within the system exhibit
delocalization for $\alpha<1$, transitioning to localization for
$\alpha>1$~\cite{PhysRevB.69.165117}. Similar to the previous models,
we set the Fermi energy at $E_F=0$. However, it is worth highlighting
that unlike the RD and AA models, the PRBA model incorporates
long-range hopping amplitudes.

We also investigate the Anderson model in one, two, and three
dimensions (1D, 2D, and 3D) with a Hamiltonian analogous to
Eq. (\ref{ham1}), featuring constant nearest-neighbor hopping
amplitudes ($t=1$). The on-site energies in this model are randomly
distributed, following a Gaussian distribution with a mean of zero and
a variance of $w$. It is well-established that in one and two
dimensions, the system becomes localized with any infinitesimal level
of disorder~\cite{markos2006numerical}, thus obviating the presence of
a delocalized-localized phase transition. However, for the 3D Anderson
model, the system remains delocalized for small values of the disorder
strength $w$, eventually transitioning to localization at a critical
value, approximately
$w_c \approx
6.1$~\cite{markos2006numerical,doi:10.7566/JPSJ.87.094703,Slevin_2014}. For
our calculations concerning the Anderson models in one, two, and three
dimensions, we maintain the Fermi energy at $E_F=0$.

\begin{figure} \centering
\includegraphics[width=0.5\textwidth]{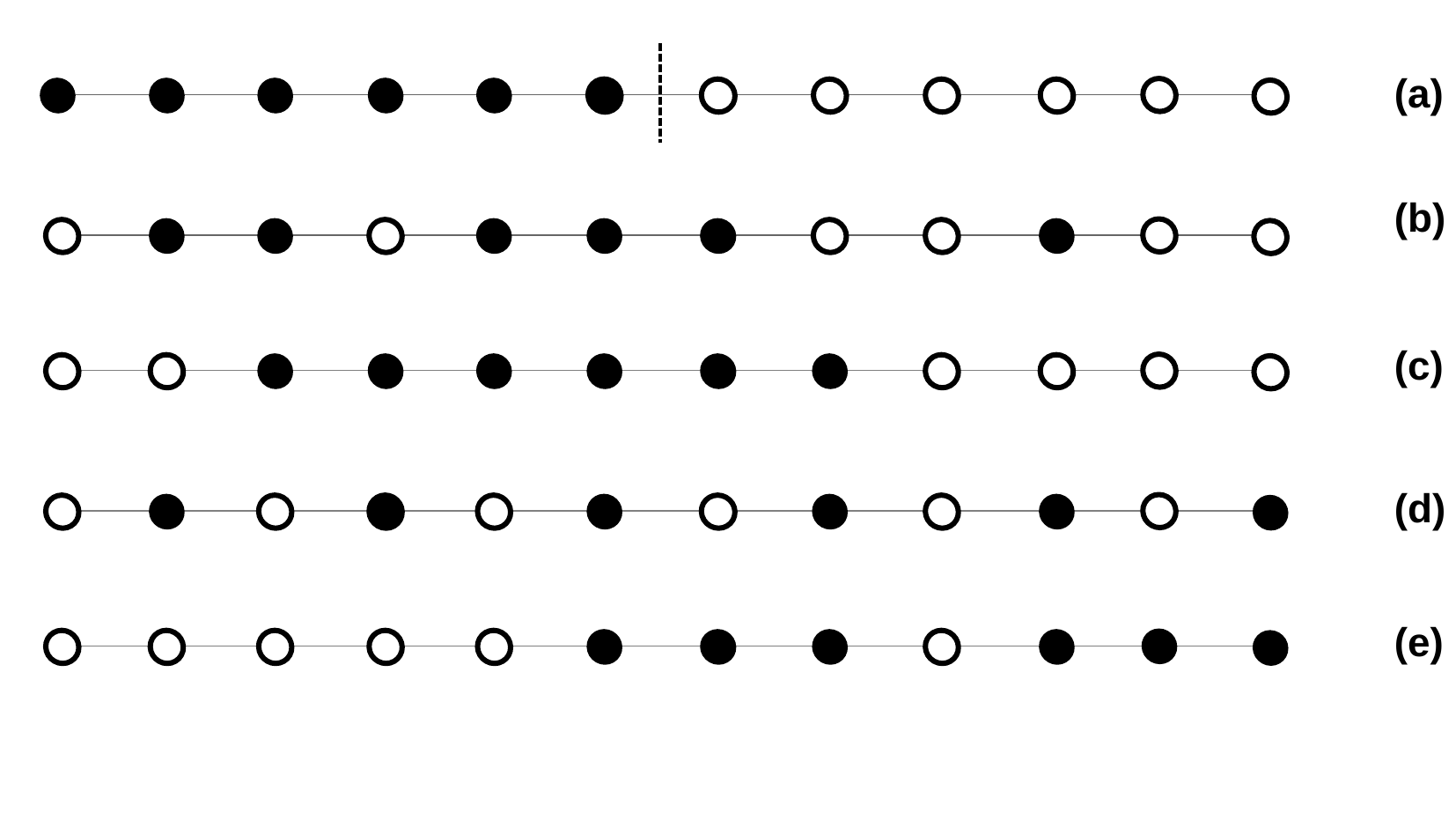}
\caption{Schematic representation of random bi-partitioning of the
  system. Sites belonging to subsystem $A$ are colored in black, while
  the white sites represent the environment. In these examples, we set
  $L=12$ and $L_A=L/2=6$. (a) depicts a typical example of
  bi-partitioning, where 6 black sites are separated from 6 white
  sites by a dashed line in the middle. (b) illustrates a typical
  example of \textit{random} bi-partitioning. (c) shows a limiting
  case where all of the subsystem sites are adjacent. (d) displays a
  limiting case with no adjacent sites within the subsystem. (e)
  demonstrates an example where the minimum number of connected sites
  is set to $\ell=3$.}
\label{fig:subsysA}
\end{figure}

To compute the entanglement entropy (EE), we first partition the
system. In the case of a lattice model comprising $L$ sites,
conventional practice involves splitting the system at its midpoint,
designating one half as subsystem $A$ (see Fig.~\ref{fig:subsysA}
(a)). However, in this report, we introduce a novel approach, namely,
\emph{random bi-partitioning}, where we randomly select $L/2$ sites
based on a uniform distribution to form subsystem $A$ (see
Fig.~\ref{fig:subsysA} (b) for a typical example of random
bi-partitioning). Notably, the sites belonging to subsystem $A$ can be
either adjacent or widely separated, resulting in a subsystem composed
of randomly distributed sites across the entire system. This
represents a departure from the conventional practice of splitting the
system at its midpoint.

To calculate the EE, we follow a practical method employing the
correlation matrix~\cite{0305-4470-36-14-101}:
\begin{equation} C_{ij} = \langle c^{\dagger}_i c_j \rangle,
\end{equation} where $i$ and $j$ traverse the indices of the randomly
chosen subsystem. EE can then be determined based on the eigenvalues
of the correlation matrix $\{\eta\}$ using the following expression:
\begin{equation} \text{EE} = -\sum_{i=1}^{L_A} \left[\eta_i \log
\eta_i + (1-\eta_i) \log (1-\eta_i)\right],
\end{equation} where $L_A=L/2$ represents the size of the subsystem.

\section{Results}\label{sec:results}

In this section, we present our detailed numerical calculations of
entanglement entropy (EE) in a randomly bi-partitioned system. We
investigate the behavior of EE in models with delocalized and
localized phases, including the random dimer (RD), Aubry-André (AA),
and power-law bond-disordered Anderson (PRBA) models. Additionally, we
explore EE in Anderson models in one, two, and three dimensions (1D,
2D, and 3D). Our primary goal is to examine the size dependence of EE
and evaluate the impact of the distribution of subsystem sites on the
EE in randomly bi-partitioned systems.

\subsection{Random Bi-partitioning in RD, AA, and PRBA Models}

\begin{figure} \centering
  \begin{subfigure}{}%
    \includegraphics[width=0.22\textwidth]{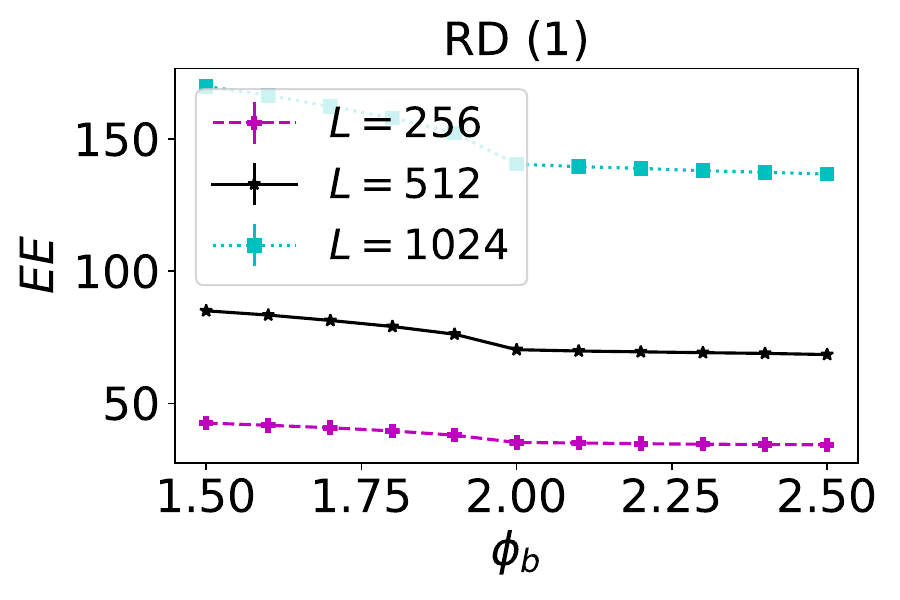}
  \end{subfigure} ~%
  \begin{subfigure}{}%
    \includegraphics[width=0.22\textwidth]{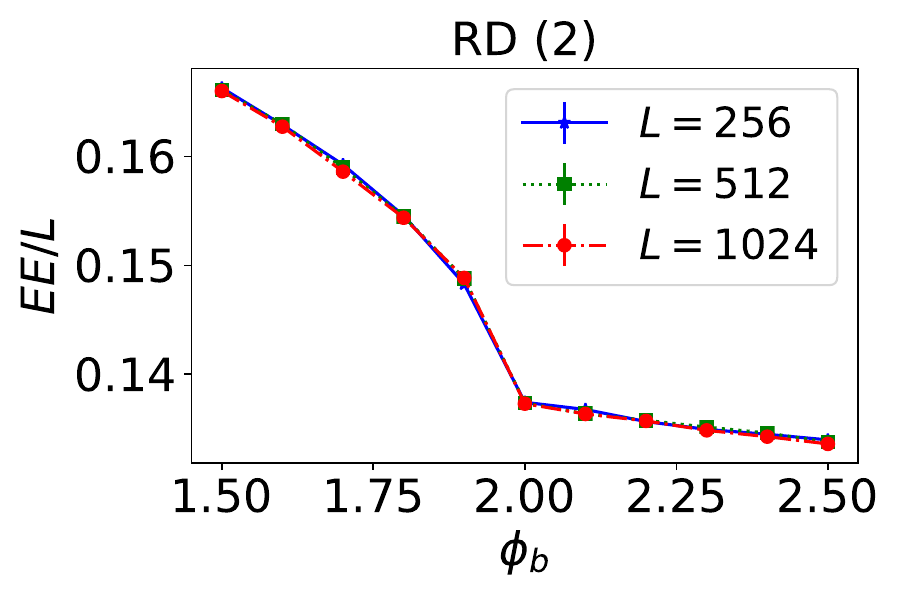}
  \end{subfigure}
  \begin{subfigure}{}%
    \includegraphics[width=0.22\textwidth]{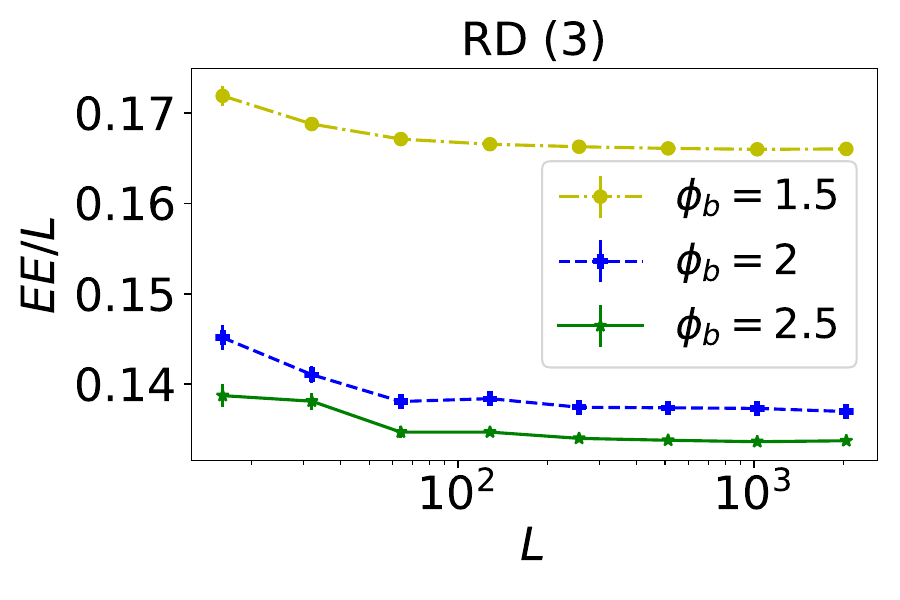}
  \end{subfigure} ~%
  \begin{subfigure}{}%
    \includegraphics[width=0.22\textwidth]{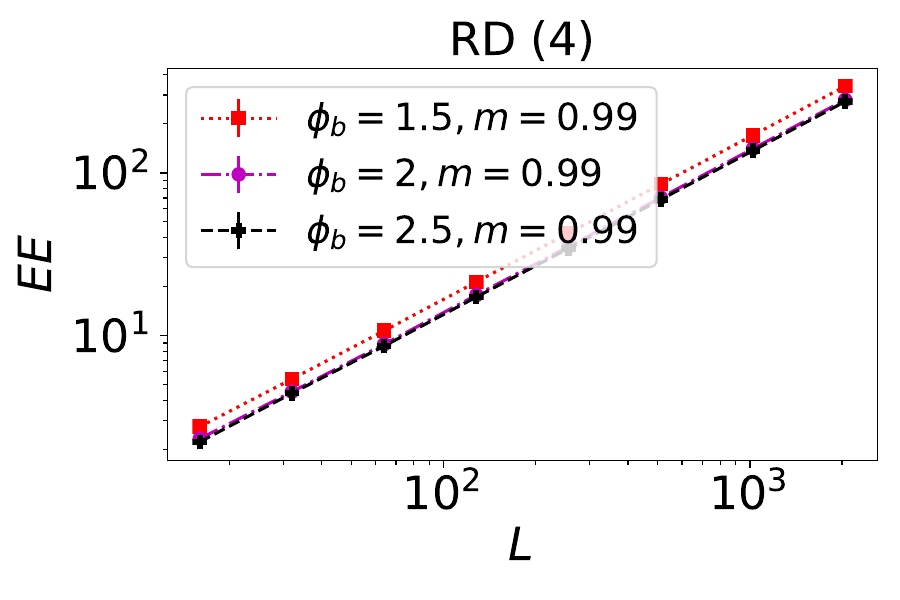}
  \end{subfigure}
	
  \caption{Entanglement Entropy (EE) Characteristics in the Random
    Dimer (RD) Model with Random Bi-partitioning.  \textbf{(1)} EE is
    depicted as a function of $\phi_b$ for various system sizes:
    $L=256, 512, 1024$.  \textbf{(2)} The behavior of $EE/L$ is shown
    for different system sizes, demonstrating a consistent trend in
    both delocalized and localized phases.  \textbf{(3)} For system
    sizes exceeding approximately $100$, $EE/L$ converges to a stable
    value contingent upon $\phi_b$.  \textbf{(4)} A power-law
    relationship between EE and system size $L$ is evident, observed
    in the log-log scale where the slope, denoted as $m$, closely
    approaches $1$. The Fermi energy is set to $E_F=\phi_b$. Each data
    point represents an average over $10^4$ samples.}
\label{fig:RD}
\end{figure}

First, we delve into the behavior of EE in a randomly bi-partitioned
system for the RD model. The results of our numerical calculations are
depicted in Fig.~\ref{fig:RD}, comprising four panels.

In Panel (1), we plot EE versus $\phi_b$ for various system
sizes. Notably, a singularity emerges in the EE at the phase
transition point, making it readily distinguishable. Furthermore, EE
in the delocalized phase surpasses that in the localized phase.

Panel (2) showcases the plot of $EE/L$ versus $\phi_b$ for different
system sizes. We observe that the behavior remains consistent across
various sizes, with the curves overlapping in both the delocalized and
localized phases.

Panel (3) presents a plot of $EE/L$ as a function of system size. It
becomes evident that $EE/L$ converges to a fixed value for large
system sizes, and this fixed value is dependent on $\phi_b$.

Finally, Panel (4) illustrates a log-log scale plot of EE versus
system size. The resulting curve exhibits a linear trend with a slope
close to 1 in both the delocalized and localized phases, indicating
that $EE$ is proportional to $L$, with the proportionality dependent
solely on $\phi_b$. Consequently, we can express $EE$ as $EE =
f_{\text{RD}}(\phi_b) L$ in both the delocalized and localized phases,
where $f$ represents a function solely dependent on $\phi_b$.

\begin{figure} \centering
  \begin{subfigure}{}%
    \includegraphics[width=0.22\textwidth]{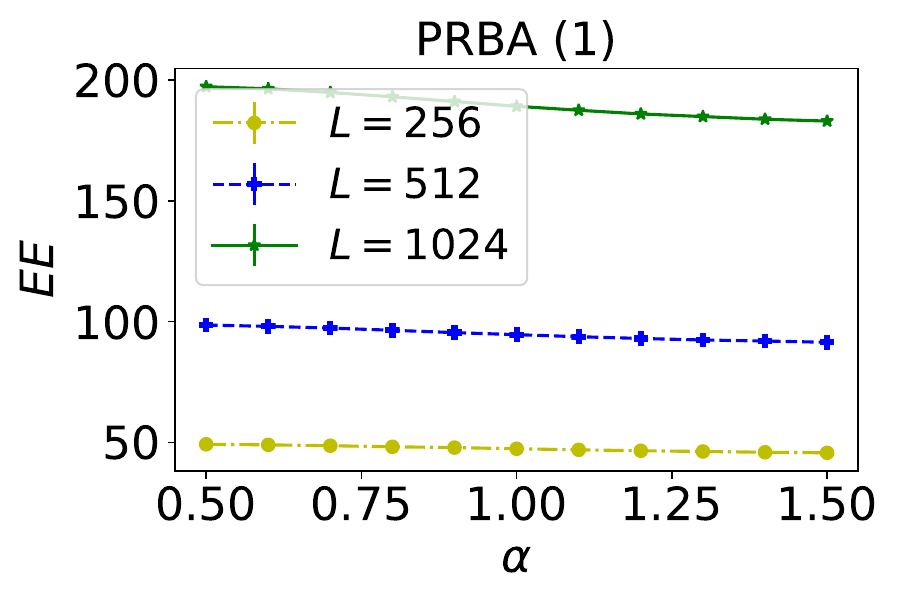}
  \end{subfigure} ~%
  \begin{subfigure}{}%
    \includegraphics[width=0.22\textwidth]{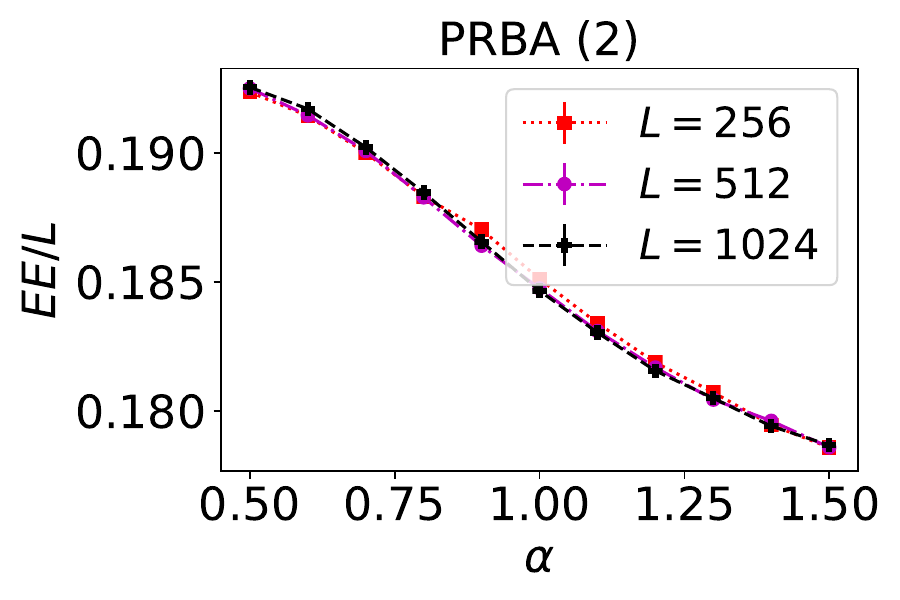}
  \end{subfigure}
  \begin{subfigure}{}%
    \includegraphics[width=0.22\textwidth]{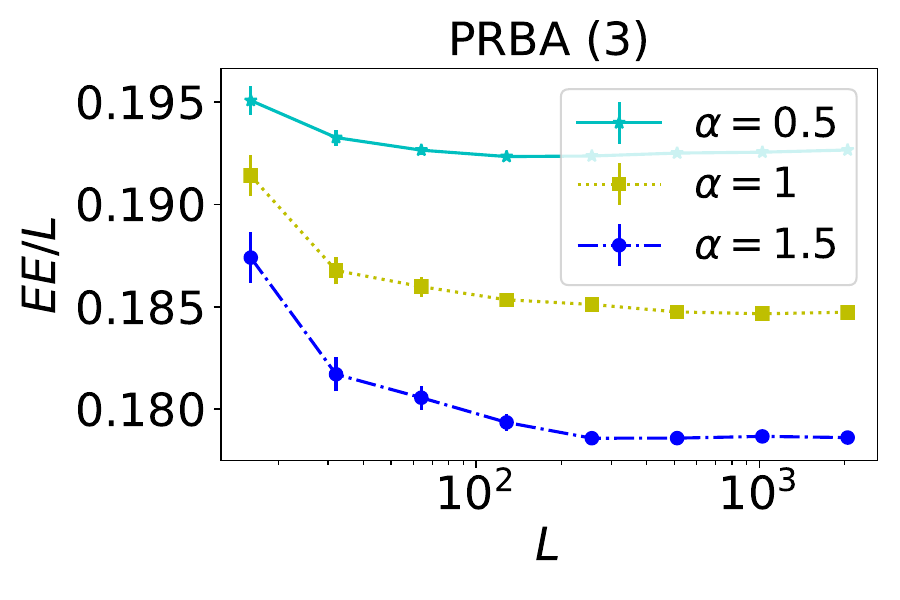}
  \end{subfigure} ~%
  \begin{subfigure}{}%
    \includegraphics[width=0.22\textwidth]{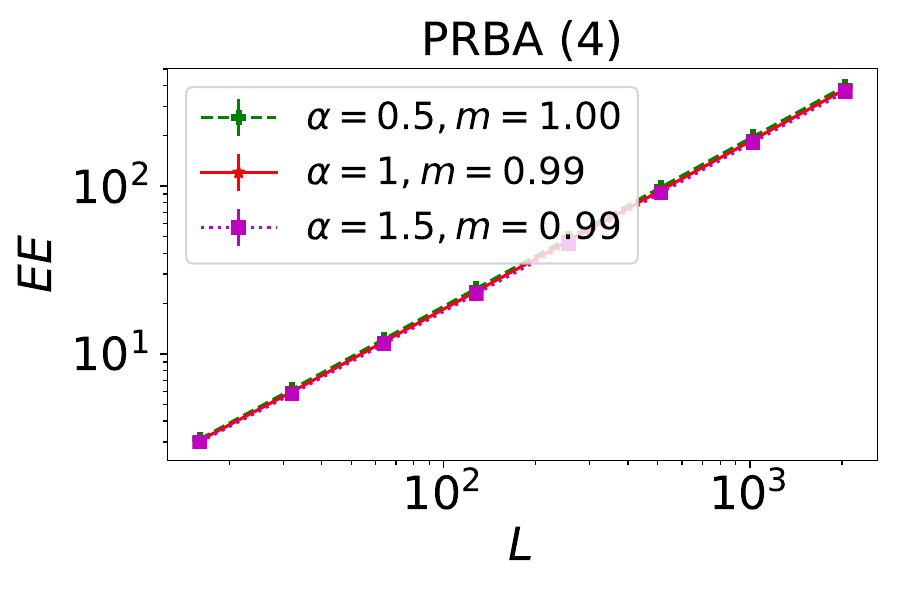}
  \end{subfigure}
	
  \caption{Analysis of the entanglement entropy (EE) behavior in the
    Power-law Bond-Disordered Anderson (PRBA) model under random
    bi-partitioning. (1) EE is depicted as a function of $\alpha$ for
    varying system sizes $L=256, 512, 1024$. (2) The trends of EE per
    site ($EE/L$) for different system sizes consistently align in
    both delocalized and localized phases. (3) EE/L reaches a steady
    state for system sizes larger than approximately $\sim 300$, with
    the steady value depending solely on $\alpha$. (4) A power-law
    relationship between EE and system size $L$ is evident (the slope
    in the log-log scale, denoted as $m$, closely approximates
    $1$). $E_F$ is held constant at $0$, and each data point is an
    average computed from $10^4$ samples. \label{fig:PRBA}}
\end{figure}

For the AA model, as shown in Fig.~\ref{fig:PRBA}, the behavior of EE
is not distinguishable at the phase transition point. Nevertheless, it
is evident that EE is smaller in the localized phase compared to the
delocalized phase (as seen in the 1st and 2nd plots). Based on the
observations from the 3rd and 4th plots, we can conclude that
$EE = f_{\text{AA}}(\lambda) L$, indicating a power-law behavior of EE
with respect to system size in both the delocalized and localized
phases.

As for the PRBA model (refer to Fig.~\ref{fig:PRBA}), we note that EE
is lower in the localized phase than in the delocalized
phase. Similarly, we can conclude that $EE = f_{\text{PRBA}}(\alpha)
L$ holds true in both the delocalized and localized phases.

\begin{figure} \centering
  \begin{subfigure}{}%
    \includegraphics[width=0.22\textwidth]{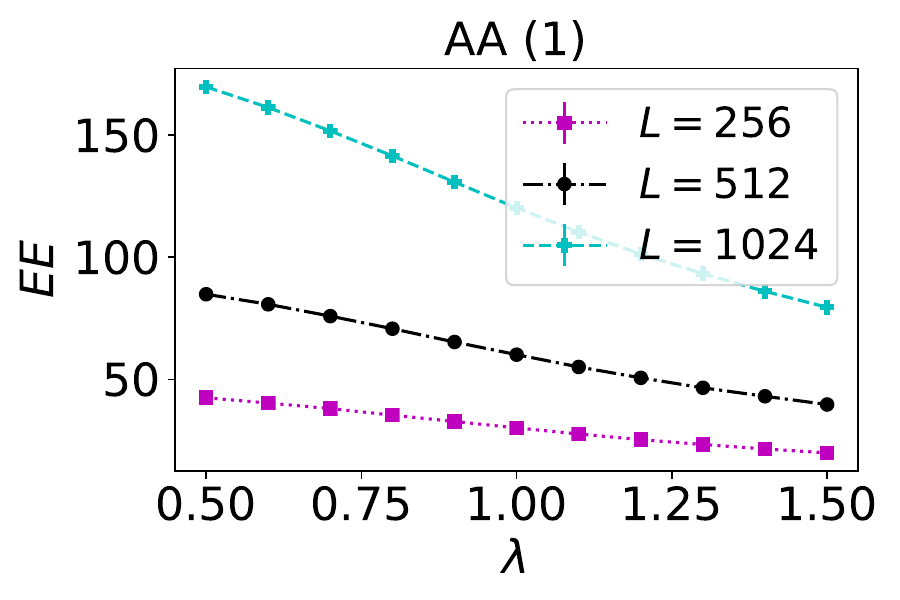}
  \end{subfigure} ~%
  \begin{subfigure}{}%
    \includegraphics[width=0.22\textwidth]{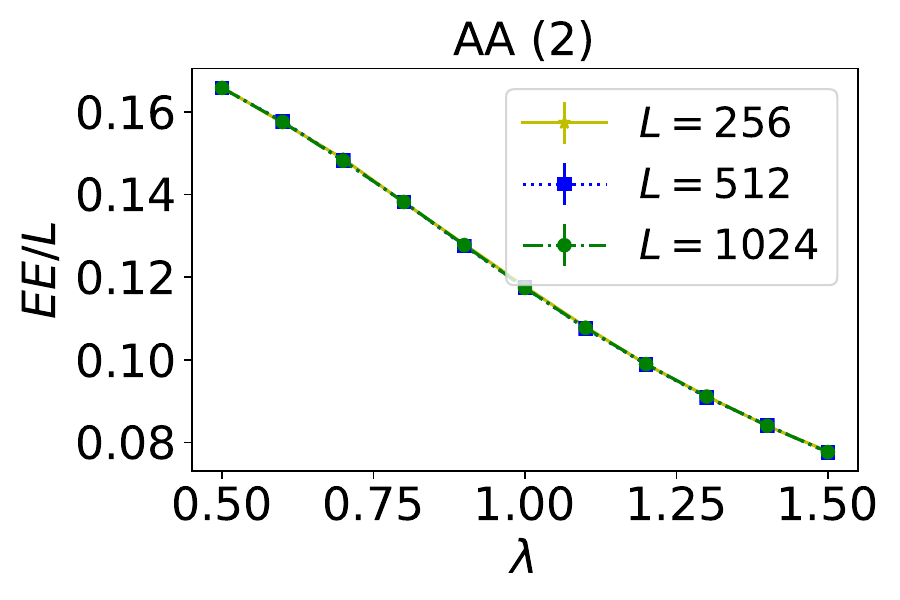}
  \end{subfigure}
  \begin{subfigure}{}%
    \includegraphics[width=0.22\textwidth]{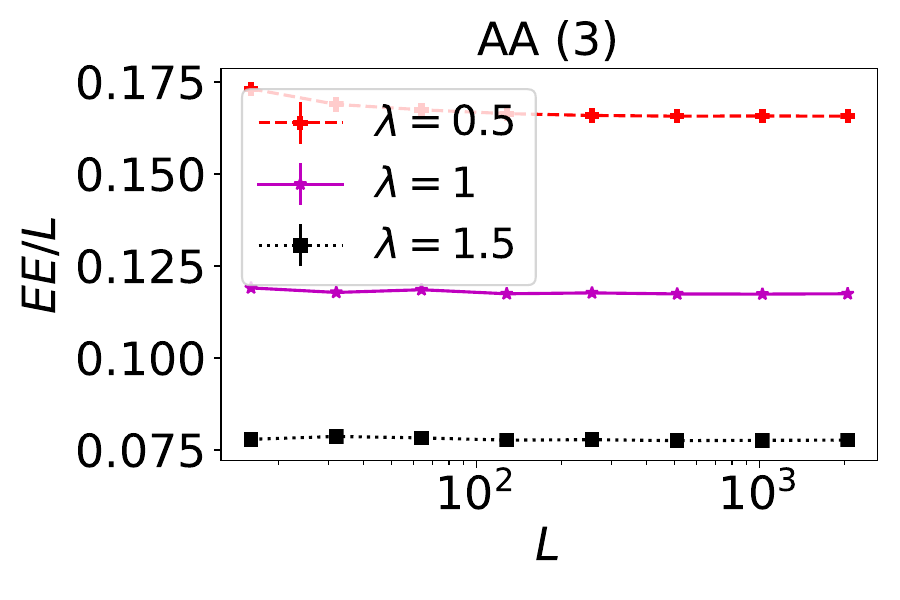}
  \end{subfigure} ~%
  \begin{subfigure}{}%
    \includegraphics[width=0.22\textwidth]{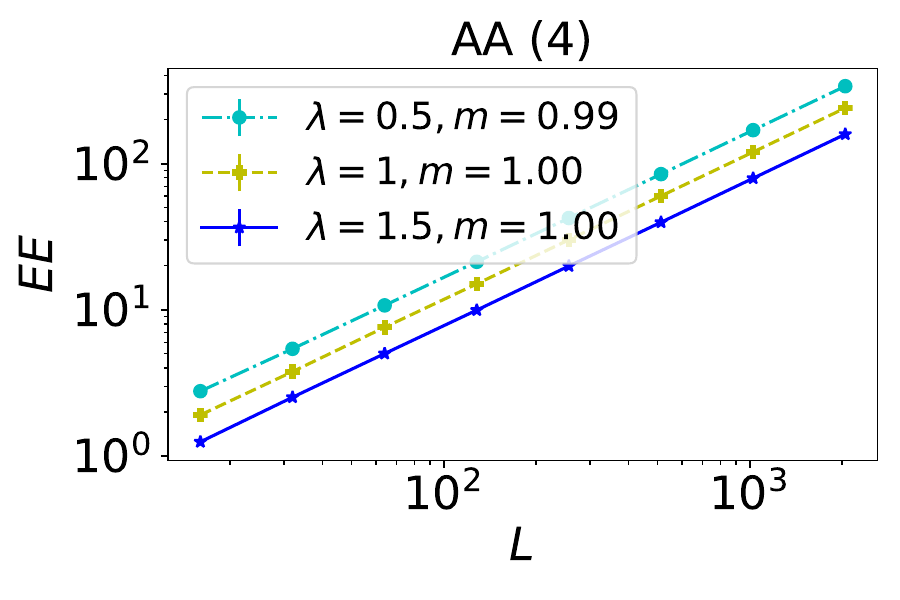}
  \end{subfigure}
	\caption{Behavior of the entanglement entropy (EE) in the
Aubry-Andre (AA) model under random bi-partitioning. (1) EE is plotted
against $\lambda$ for various system sizes $L=256, 512, 1024$. (2) The
behaviors of EE per site ($EE/L$) for different system sizes
consistently coincide in both delocalized and localized phases. (3)
EE/L saturates to a constant value for system sizes beyond
approximately $\sim 100$, depending solely on $\lambda$. (4) An
observed power-law behavior of EE versus system size $L$ (where the
slope in the log-log scale, denoted as $m$, closely approaches
$1$). $E_F$ is set to $0$, and each data point results from averaging
over $10^4$ samples.
\label{fig:AA}}
\end{figure}

\subsection{Random Bi-partitioning in the Anderson Model in
One, Two, and Three Dimensions}
	
\begin{figure*} \centering
  \begin{subfigure}{}%
    \includegraphics[width=0.32\textwidth]{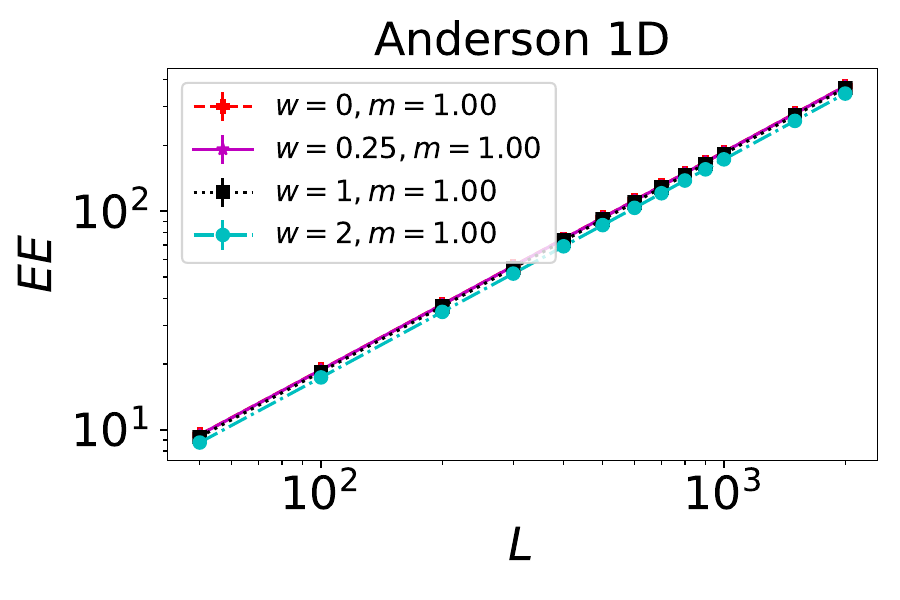}
  \end{subfigure}
  \begin{subfigure}{}%
    \includegraphics[width=0.32\textwidth]{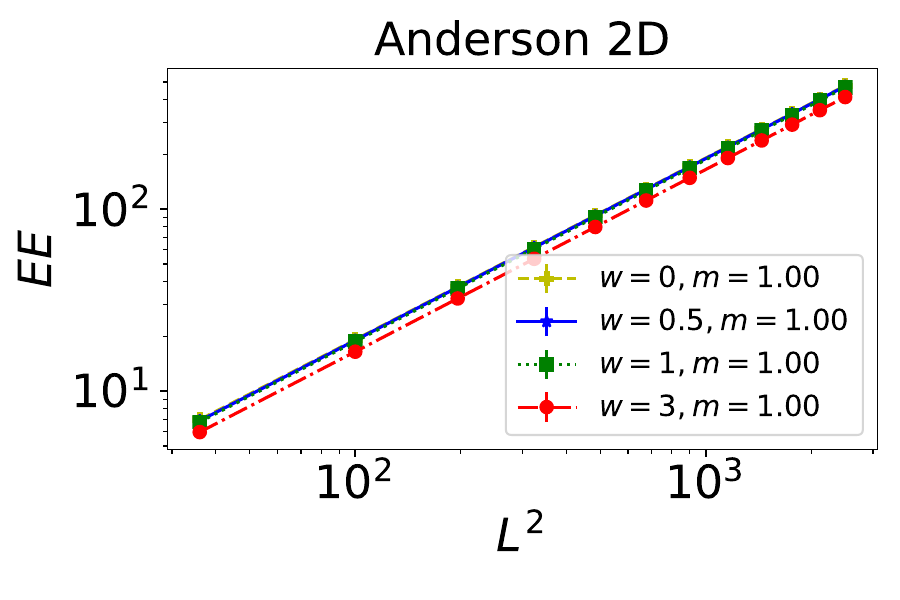}
  \end{subfigure}%
  \begin{subfigure}{}%
    \includegraphics[width=0.32\textwidth]{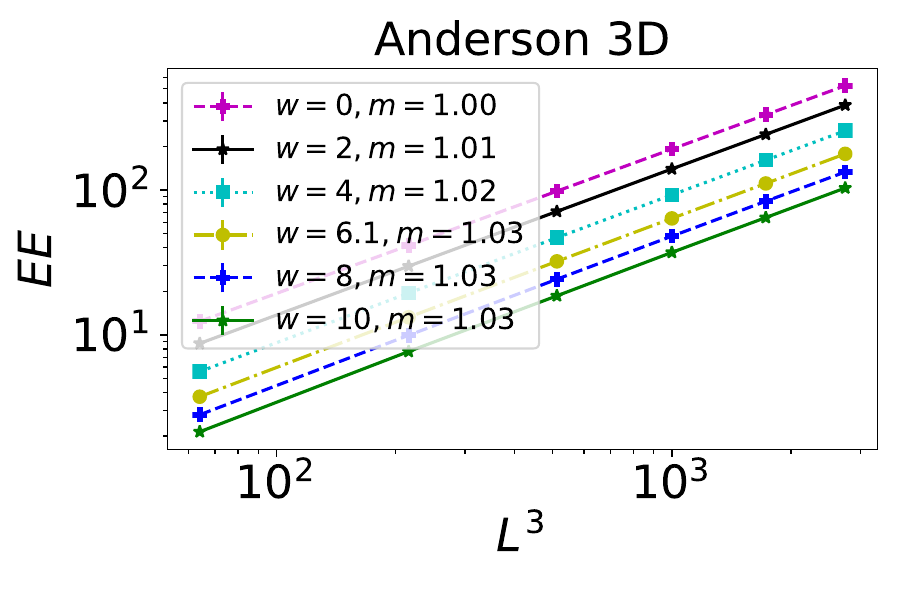}
  \end{subfigure}%
  \caption{Scaling behavior of the entanglement entropy ($EE$) with
    respect to the system volume for the Anderson models in one
    dimension (1D), two dimensions (2D), and three dimensions (3D),
    depicted in the left, middle, and right panels, respectively. The
    log-log plots exhibit power-law scaling, with the slope ($m$) of
    the lines consistently approximating unity ($1$). $E_F$ is set to
    $0$ for all cases, and each data point is obtained by averaging
    over $10^4$ samples for the 1D model, $10^3$ samples for the 2D
    model, and $10^2$ samples for the 3D model. \label{fig:anderson}}
\end{figure*}

In this subsection, we undertake an examination of the entanglement
entropy (EE) with the application of random bi-partitioning within the
Anderson model across one, two, and three dimensions. It is
well-established that a phase transition between delocalized and
localized states exclusively manifests in the three-dimensional
Anderson model, whereas all states in one and two dimensions become
localized even with infinitesimal
disorder~\cite{markos2006numerical}. Our numerical findings are
presented in Figure~\ref{fig:anderson}.
	
The behavior of the EE in the one-dimensional (1D), two-dimensional
(2D), and three-dimensional (3D) Anderson models reveals a
characteristic power-law relationship. As illustrated in
Figure~\ref{fig:anderson}, the log-log plots' slopes are consistently
close to unity ($1$). This observation leads us to the conclusion
that, in the Anderson model, the entanglement entropy ($EE$) scales as
$L^D$ for both the delocalized and localized phases.

\subsection{Connected and Disconnected Subsystems}
	
In the aforementioned calculations, which employed random
bi-partitioning, it is crucial to note that the sites constituting the
subsystem were selected at random, following a uniform
distribution. Consequently, the subsystem may consist of either
connected or disconnected sites, representing two distinct
scenarios. These scenarios are exemplified by a subsystem composed of
$L_A$ connected sites (as depicted in Fig.~\ref{fig:subsysA} (c)) and
a subsystem comprising $L_A$ disconnected sites (as shown in
Fig.~\ref{fig:subsysA} (d)).
	
When assessing the Entanglement Entropy (EE), we are essentially
indirectly quantifying the correlations between the subsystem's sites
and the remainder of the system. These correlations can exhibit either
short-range or long-range behavior. Short-range correlations give rise
to entanglement only when two sites belonging to different subsystems
are situated in close proximity to the boundary. In contrast,
long-range correlations can lead to entanglement even when the
correlated sites are widely separated from one another.
	
In our investigations, the choice of subsystem composition within the
random bi-partitioning method plays a pivotal role in the resulting
entanglement entropy (EE). The subsystem may consist of completely
disconnected sites, leading to the measurement of both short-range and
long-range correlations across the entire system. Even in scenarios
where the system lacks long-range correlations, a substantial EE
emerges due to the inclusion of all short-range
correlations. Conversely, for a fully connected subsystem, only
short-range correlations in proximity to the boundary contribute to
the entanglement, alongside long-range correlations.
	
\begin{figure*} \centering
  \begin{subfigure}{}%
    \includegraphics[width=0.32\textwidth]{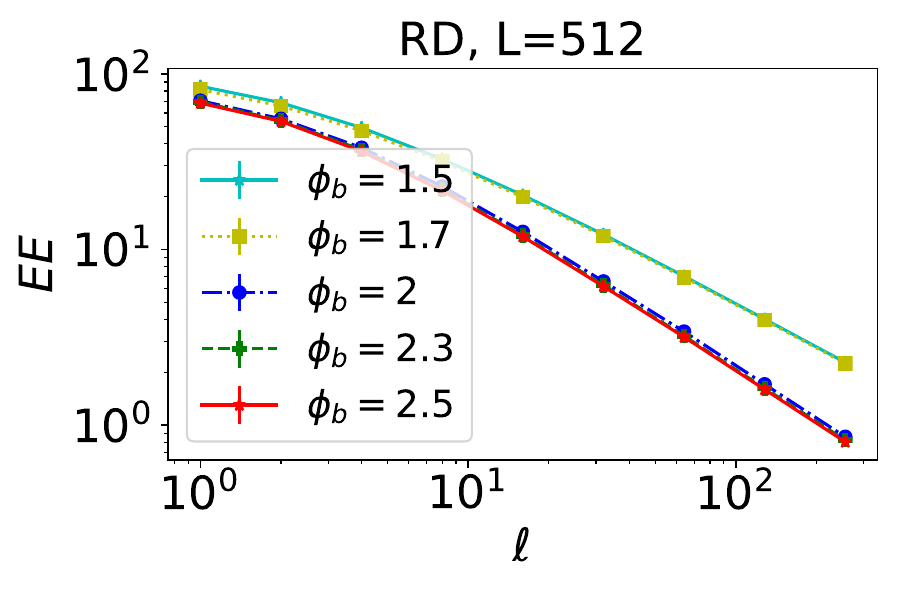}
  \end{subfigure} ~%
  \begin{subfigure}{}%
    \includegraphics[width=0.32\textwidth]{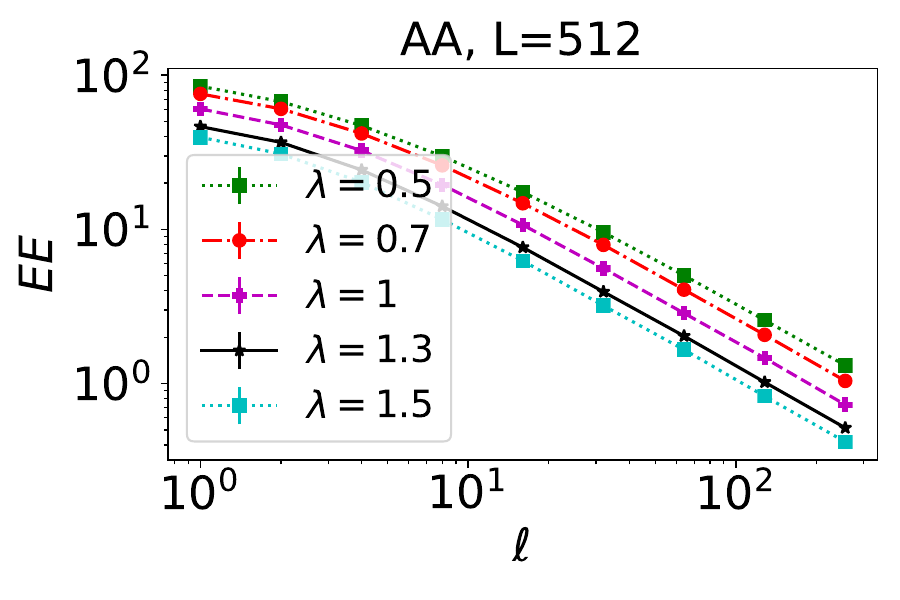}
  \end{subfigure}
  \begin{subfigure}{}%
    \includegraphics[width=0.32\textwidth]{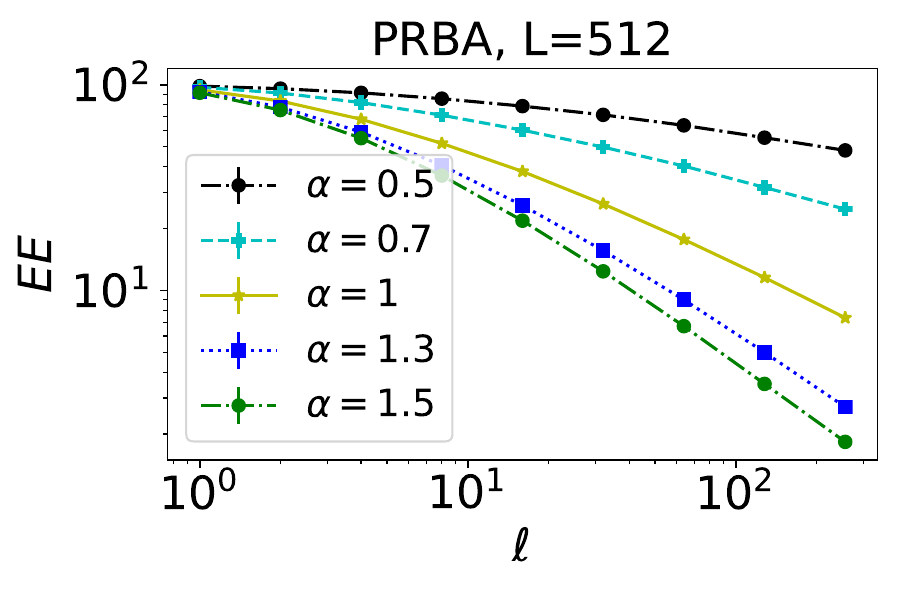}
  \end{subfigure}
  \caption{Variation of the Entanglement Entropy (EE) with the minimum
    number of connected sites, denoted as $\ell$, ranging from $1$ to
    $L/2=256$ for a system of size $L=512$ in the Random Disordered
    (RD) model (left panel), Aubrey-André (AA) model (middle panel),
    and Power-Law Bond-Disordered Anderson (PRBA) model (right
    panel). The EE exhibits a decreasing trend as $\ell$ increases,
    with this decrease being more pronounced in the localized phase
    compared to the delocalized phase. Each data point is obtained by
    averaging over $10^4$ samples (see Table~\ref{t:ell}).
\label{fig:ell}}
\end{figure*}

This observation underscores the significance of the distribution of
subsystem sites in EE calculations. To elucidate this point, we
introduce a parameter, $\ell$, which represents the minimum number of
connected sites within the subsystem (Fig.~\ref{fig:subsysA} (e)
depicts the case where $\ell=3$). The value of $\ell$ varies within
the range of $1$ to $L/2$. In Fig.~\ref{fig:ell}, we present the
variation of $EE/L$ with $\ell$ for systems of size $L$, considering
the RD, AA, and PRBA models.
	
As demonstrated in Fig.~\ref{fig:ell}, an increase in the parameter
$\ell$ results in a corresponding decrease in entanglement. This
behavior is consistent with expectations, as a higher value of $\ell$
diminishes the contributions to EE stemming from short-range
correlations, leading to a reduction in EE.
		
However, the decrease in the EE is not uniform across all considered
models. To quantify these variations, we introduce a measure that
characterizes the change in $\log_{10} EE$ as we vary $\ell$ from $1$
to $L/2$ in both delocalized and localized phases:
	
\begin{equation}
  \Delta_{D/L} = \frac{\log_{10}[EE_{\ell=1}] -
    \log_{10}[EE_{\ell=L/2}]}{\log_{10}[EE_{\ell=1}]},
\end{equation}
	
where $\Delta_D$ ($\Delta_L$) represents the ratio of change in the
delocalized (localized) phase. Specifically, for the delocalized
phase, we set $\phi_b=1.5$ for the RD model, $\lambda=0.5$ for the AA
model, and $\alpha=0.5$ for the PRBA model. In contrast, for the
localized phase, we choose $\phi_b=2.5$ for the RD model,
$\lambda=1.5$ for the AA model, and $\alpha=1.5$ for the PRBA model,
considering parameter values deep within each respective phase. The
numerical results are tabulated in Table \ref{t:ell}.
	
\begin{table}
  \begin{ruledtabular}
    \caption{Table of the ratio of change in $\log_{10}EE$ when
      increasing the minimum number of connected sites, $\ell$, from
      $1$ to $L/2=256$ in the delocalized phase $\Delta_D$ and in the
      localized phase $\Delta_L$, for RD, AA, and PRBA models, as per
      the data in Fig.~\ref{fig:ell}. \label{t:ell}}
    \begin{tabular}{c|c|c}
      Model & $\Delta_D$ & $\Delta_L$\\
      \hline
      RD & 0.81 & 1.05 \\
      AA & 0.93 & 1.23 \\
      PRBA & 0.15 & 0.86 \\
    \end{tabular}
  \end{ruledtabular}
\end{table}
	
To summarize our observations, we note the following:
	
Firstly, we consistently observe that the decrease in the entanglement
entropy (EE) is more pronounced in the localized phase compared to the
delocalized phase, i.e., $\Delta_L > \Delta_D$. This phenomenon can be
attributed to the presence of both short and long-range correlations
in the delocalized phase. In contrast, the localized phase
predominantly exhibits short-range correlations. Consequently,
increasing $\ell$ in the localized phase results in the omission of a
significant portion of short-range correlations, leading to a more
substantial reduction in EE.
	
Secondly, it is worth highlighting that $\Delta_D$ for the PRBA model
is smaller than that of the RD and AA models. This difference can be
attributed to the distinctive hopping amplitudes in the PRBA
model. Specifically, the PRBA model features long-range hopping
amplitudes, which differ from the RD and AA models that primarily
involve nearest-neighbor hopping amplitudes. Consequently, the PRBA
model exhibits longer-range correlations, resulting in a relatively
smaller decrease in the entanglement entropy in the delocalized phase
compared to the RD and AA models, which primarily rely on short-range
hopping amplitudes.
	
These observations shed light on the intricate interplay between
correlation lengths, hopping amplitudes, and the behavior of
entanglement entropy in diverse phases of the studied models.
	
\begin{figure*} \centering
  \begin{subfigure}{}%
    \includegraphics[width=0.32\textwidth]{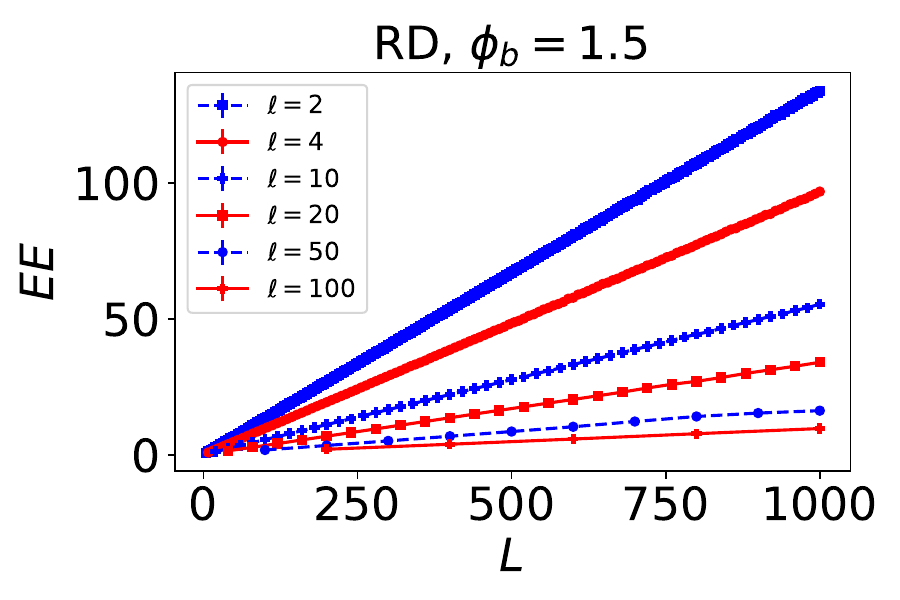}
  \end{subfigure} ~%
  \begin{subfigure}{}%
    \includegraphics[width=0.32\textwidth]{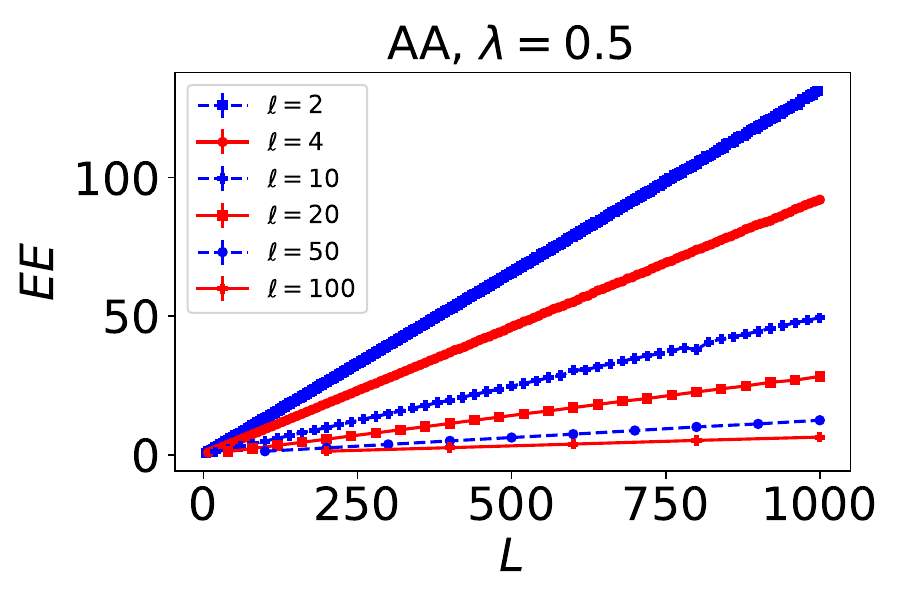}
  \end{subfigure}
  \begin{subfigure}{}%
    \includegraphics[width=0.32\textwidth]{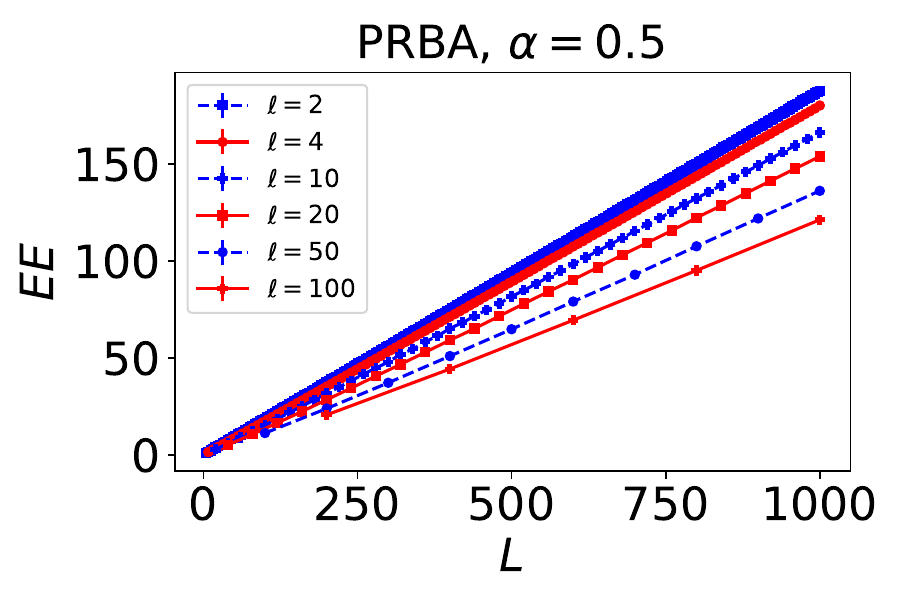}
  \end{subfigure} ~%
  \begin{subfigure}{}%
    \includegraphics[width=0.32\textwidth]{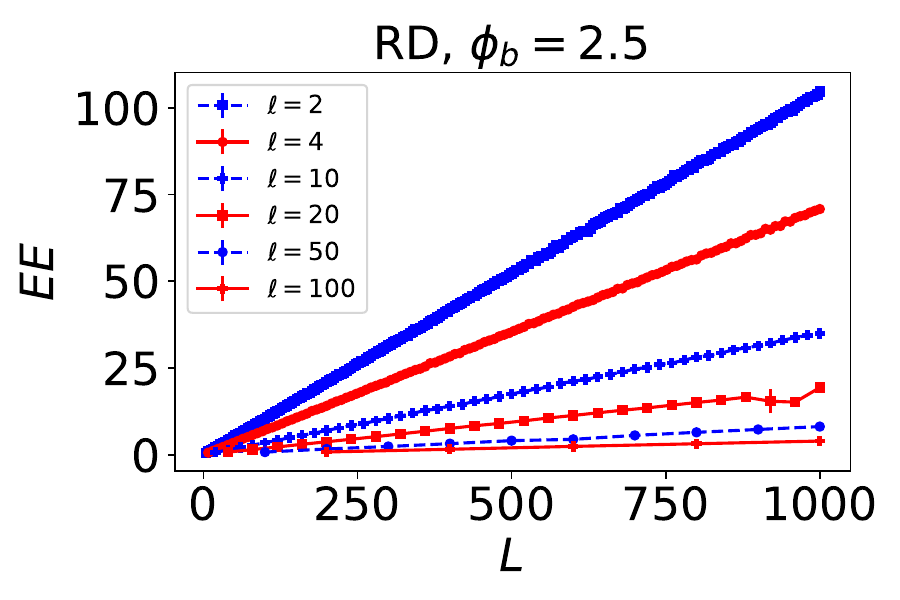}
  \end{subfigure} ~%
  \begin{subfigure}{}%
    \includegraphics[width=0.32\textwidth]{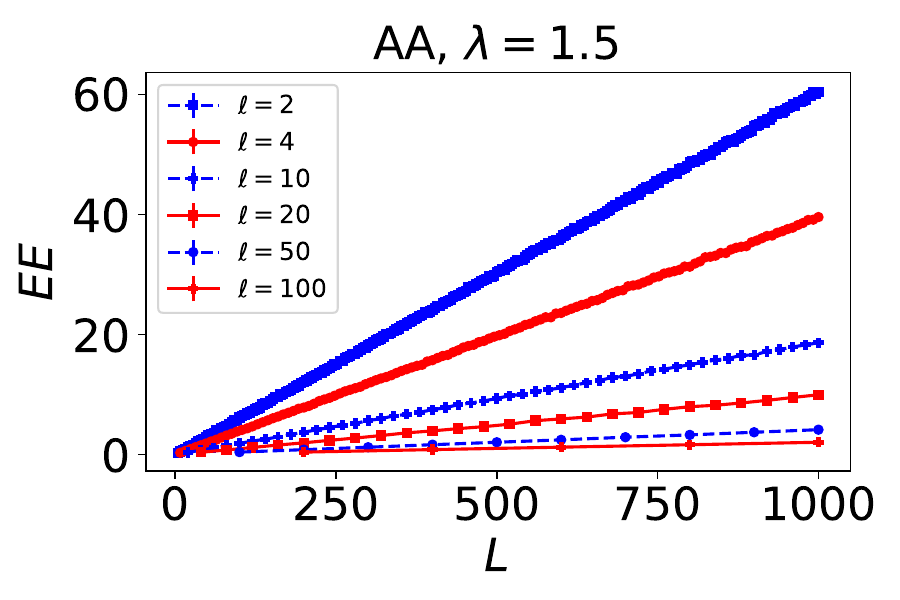}
  \end{subfigure} ~%
  \begin{subfigure}{}%
    \includegraphics[width=0.32\textwidth]{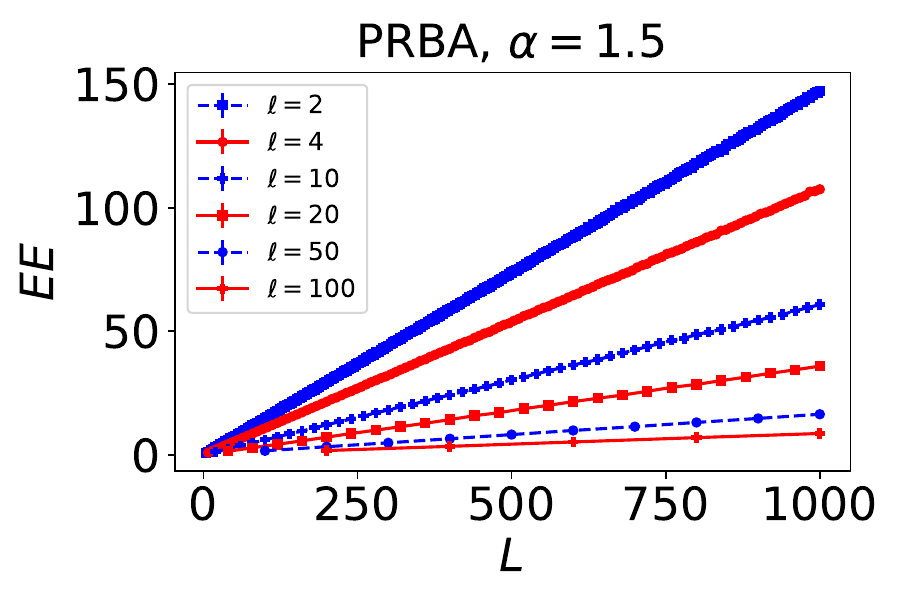}
  \end{subfigure}
  \caption{Entanglement entropy (EE) versus system size
$L$ is shown in the top panels for the delocalized phase and in the
bottom panels for the localized phase. The plots depict the behavior
of the RD model (left panels), AA model (middle panels), and PRBA
model (right panel). An increase in the parameter $\ell$ leads to a
crossover from a volume-law to an area-law scaling. Each data point
represents the average over $10^4$ samples.
\label{fig:EEvsL_ell}}
\end{figure*}

In our analysis of the entanglement entropy (EE) as a function of the
system size $L$ for various values of $\ell$, presented in
Fig. \ref{fig:EEvsL_ell}, we made the following observations:
	
i) As expected, the EE consistently exhibits lower values in the
localized phase when compared to the delocalized phase, across all
considered values of $\ell$ and system sizes.
	
ii) We observed a notable transition from a volume-law to an area-law
scaling behavior as $\ell$ is increased. For smaller $\ell$ values,
the EE displays rapid linear growth concerning the system size,
adhering to a volume-law scaling. Conversely, as $\ell$ is increased,
the rate of EE growth diminishes, resulting in a saturation behavior
that conforms to an area-law scaling.
	
iii) It is noteworthy that the saturation point occurs at smaller
values of $\ell$ in the localized phase compared to the delocalized
phase.
	
iv) The PRBA model, characterized by long-range hopping amplitudes,
exhibits distinct behavior. In the delocalized phase, the rate of EE
increase concerning the system size is notably higher for all choices
of $\ell$ compared to the other models.
	
These findings provide valuable insights into the intricate scaling
behaviors of EE, which depend on the phase of the system, the
subsystem size, and the range of hopping amplitudes present in the
model.

\section{Conclusion}\label{sec:outlook}
	
In this study, we have examined the entanglement properties of
disordered free fermion systems by employing random
bi-partitioning. The calculation of entanglement entropy (EE) involves
dividing the system into two subsystems, typically achieved by cutting
the system at the middle and considering the first half as the
subsystem. However, in this report, we have adopted an alternative
approach by randomly selecting sites from the entire system to form
the subsystem. This unconventional subsystem configuration has
implications for the resulting entanglement properties.
	
Our analysis focused on the behavior of EE in free fermion models with
delocalized and localized phases, specifically utilizing Anderson
models in one, two, and three dimensions. We have found that the
behavior of EE remains smooth across the phase transition point,
consistent with our previous conclusions
(Ref. \cite{POURANVARI2023128908}) that EE with random partitioning
captures both long-range and short-range correlations throughout the
system. Consequently, as we traverse the phase transition from the
delocalized phase to the localized phase, the long-range correlations
decrease, but short-range correlations persist throughout the system.
	
Furthermore, we have observed that EE increases with increasing system
size, exhibiting a power-law scaling with system size ($L$) in $D$
dimensions: $EE \propto L^D$ in both the delocalized and localized
phases. These findings are based on detailed numerical calculations
performed on free fermion models, including Anderson models in one,
two, and three dimensions.
	
Additionally, we have examined the influence of subsystem site
distribution on EE and noted that having more adjacent sites belonging
to the subsystem leads to a decrease in EE. This observation can be
attributed to the indirect measurement of correlations by EE,
encompassing both short-range and long-range correlations. Increasing
the number of connected sites in the subsystem results in the loss of
information regarding short-range correlations. Moreover, in the
delocalized phase, when significant long-distance hopping amplitudes
exist in the Hamiltonian, EE demonstrates greater robustness to
changes in site distribution.
	
It is worth mentioning that the aforementioned random partitioning can
also be applied to the partitioning in momentum space
(Ref. \cite{PhysRevLett.110.046806}), where the distribution of
subsystem sites directly impacts occupied and unoccupied levels.
	
In conclusion, this study represents one of the initial investigations
into the entanglement properties of disordered free fermion systems
exhibiting delocalized-localized phase transitions. As such, it
contributes to the growing body of research exploring the behavior of
entanglement entropy in these systems. The use of random
bi-partitioning provides a unique approach to examining the interplay
between subsystem configurations, phase transitions, system
dimensions, and correlation effects. These findings hold significance
for the physics community, shedding light on the intricate nature of
entanglement in disordered fermion systems and their transition
between delocalized and localized phases.

	\acknowledgments
	
	The author gratefully acknowledge the high performance
computing center of university of Mazandaran for providing computing
resource and time.
	
	\bibliographystyle{apsrev4-2.bst}
\bibliography{/home/cms/Dropbox/physics/Bib/reference.bib}
\end{document}